# NViz– A General Purpose Visualization tool for Wireless Sensor Networks

Anh-Vu Dinh-Duc
HCMC University of Technology
Ho Chi Minh city, Vietnam
anhvu@cse.hcmut.edu.vn

The-Hien Dang-Ha
HCMC University of Technology
Ho Chi Minh city, Vietnam
thehien.dangha@gmail.com

Ngoc-An Lam
HCMC University of Technology
Ho Chi Minh city, Vietnam
ngocan1989@gmail.com

*Abstract*—In a Wireless Sensor Network (WSN), data manipulation and representation is a crucial part and can take a lot of time to be developed from scratch. Although various visualization tools have been created for certain projects so far, these tools can only be used in certain scenarios, due to their hard-coded packet formats and network's properties. To speed up development process, a visualization tool which can adapt to any kind of WSN is essentially necessary. For this purpose, a general-purpose visualization tool – NViz, which can represent and visualize data for any kind of WSN, is proposed. NViz allows user to set their network's properties and packet formats through XML files. Based on properties defined, user can choose the meaning of them and let NViz represents the data respectively. Furthermore, a better Replay mechanism, which lets researchers and developers debug their WSN easily, is also integrated in this tool. NViz is designed based on a layered architecture which allows for clear and well-defined interrelationships and interfaces between each components.

## I. INTRODUCTION

Recently, there has been increasing interest in WSNs that are constructed by sensor nodes. Many WSN applications in the fields of environmental monitoring, security, industrial control, and object tracking have been introduced and deployed in practice. Despite the rapid development of sensor networks, there are still many challenges developers face in deployment of WSN applications:

*1) Information visualizations:* Visualization tools are used to gather information about network properties such as: node status (e.g., battery level or communication power), network topology, wireless bandwidth, or link state… and present them in graphs or other forms on screen. However, WSNs are always application-specific which require different network properties to be monitored. Besides, different packet formats are also needed to be designed for different application scenarios. Therefore, most existing visualization tools today are developed under specialized applications which are almost useless when applications changed.

*2) Unit conversion:* Raw sensor data collected from sensor nodes are typically provided as direct digital readings from an ADC (Analog to Digital Converter). Voltage of the sensor is sampled and converted to a number relative to some reference voltages. To make the data useful, visualization tools should provide the ability of converting raw sensor data to engineering units based on conversionfunction defined by user. Most existing visualization tools are lack of that ability.

*3) Replay:* The rapidly changing state of sensor network due to fast incoming packets makes it hard for users to debug their WSNs. Researchers and developers always want to observe the detail variations inside their network to verify their algorithms or to validate their designs. However, most existing visualization tools only provide a basic replay function which helps users record their incoming packet in log files and playback them later. Users can not determine when to start playing back or run forward/backward through the log files. A new "record and play back" mechanism to enable user to monitor their WSNs like videos is therefore needed.

In this paper, we present NViz, a general purpose visualization tool which is designed to take these challenges into account. Users now can set up their network properties and packet formats through 2 XML files. Moreover, conversion function for each network properties can also be defined in XML file.

## II. RELATED WORKS

In recent years, there are many visualization tools have been introduced such as: SpyGlass[3], MoteView[4], NetTopo[5], Octopus[6] … However, most of them are developed for specialized applications and cannot verify with the change of application scenarios[1, 2]. Users have to modify and re-compile program or create new plug-in module for these tools if they want to use it in their applications.

NetViewer[7], a universal visualization tool for WSN, has been introduced at GLOBECOM 2010. It allows users define packet format in XML file and translates packets according to that. In that way, the development of WSNs and visualization tool could be separated. The Netviewer can be used for visualization no matter what the node programs are or which the protocols the base station node uses to communicate with the host. It also provides replay function for debugging purpose and on-line data dissemination for data sharing.

However, NetViewer accepts only one packet format, limiting it to be used as a general purpose visualization tool. In practice, WSNs can use many packet formats for different purposes such as association notification, event update or data update. Moreover, replay function of NetViewer is also useless in a large-scale WSN because it only supports users to play, step by step, from beginning through the data log file.

## III. DESIGN OF NVIZ

NViz is designed and divided into server and client divisions (Fig. 1) where the server part handles packet access and all information storage. It also represents network behaviors in real time. The client part allows user to remote access to database and review historical network behaviors like a video thank to checkpoint mechanism in server-side.

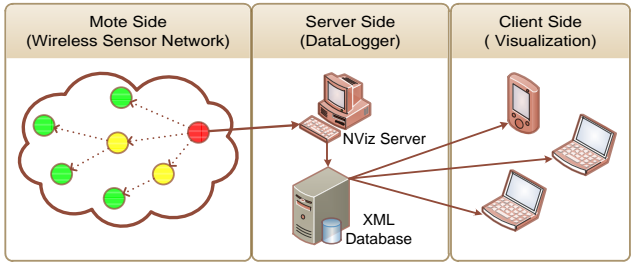

Figure 1. Client-Server design

### A. Architecture of NViz – Server division

In server-side, NViz is designed and divided into five independent layers and a cross-layer component (Fig.2) which can be developed and tested separately.

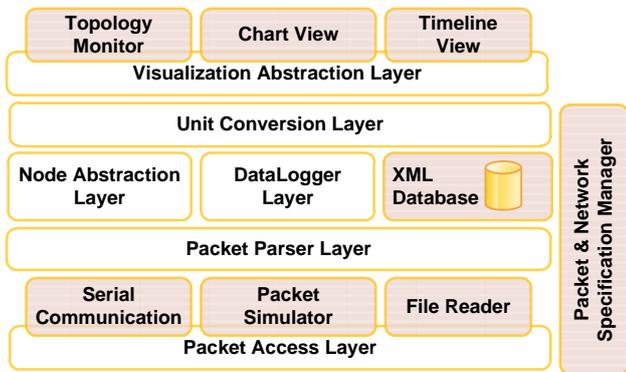

Figure 2. Architecture of NViz – Server division

*Packet and Network Specification Manager* is a component involved in four lower layers. By loading two XML specification files, NViz allows users to configure this component and lets it provide information about *Network properties* and *Packet formats* to other layers (see Section IV).

Furthermore, each layer in the architecture takes responsibility for handling one part of visualization and can be modified without any further influence on other layers.

*1) Packet Access Layer:* This layer is responsible for generating packets. Currently, we have developed three approaches for packet access. Packets can be received from a sink node that connected to host, generated by user through Packet Simulator component or read from an input file. NViz currently supports TinyOS serial communication library. Moreover, with file access approach, NViz can also be used to support other simulator programs like ns-2 or OMNeT++.

*2) Packet Parser Layer:* this layer is reponsible for parsing valid packets received and discarding the bad packets. Packets will be parsed due to network properties and packet formats information contained in two XML files. These files are defined and loaded to Packet & Network Specification Manager component by user at startup.

*3) Node Abstraction* and *Datalogger Layer:* By using network properties information provided by Packet & Network Specification Manager component, Node Abstraction component will abstracts virtual nodes, links and enviroment. Information received from Packet Parser layer will be cached in this layer and can be accessed easily by upper layer as node property, link property and enviroment property. Dataloger component in this layer is responsible for generating checkpoints and logs coresponding to current network state presented by Node Abstraction component (see Section VI).

*4) Unit Conversation Layer:* In this layer, all property values from *Node Abstration* component will be converted by coresponding conversion function (see Section V).

*5) Visualization Abstraction Layer:* This layer provides interface for user to interact with the data collected. Many visualizating services can be easily implemented in this layer by using data services provided in lower layer. Currently, this layer consists of three component: *Topology Monitor*, *Chart View* and *Timeline View* component.

### B. Architecture of NViz – Client division

Components within NViz client are conceptually split into one of five layers and a cross-layer component. Among them, three upper layers take the same responsibility as in server division. In addition, *Packet & Network Specification Manager* component will be initialized by loading two XML files stored in server's XML database.

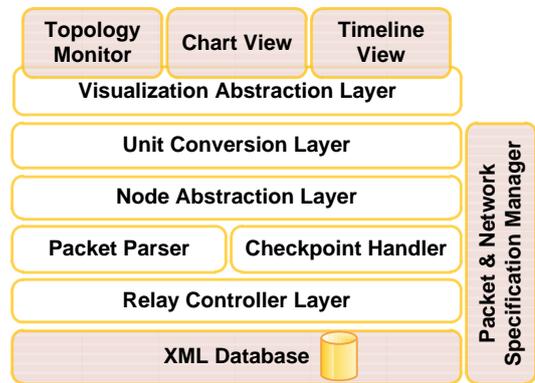

Figure 3. Architecture of NViz – Client division

Other two layers' responsibility is described below:

*1) Relay Controller Layer:* This layer uses a new machenism to allow user review their historical networks

activities easily and efficiently. Depending on user's requests, this layer will starts a system timer to generate packets from checkpoints and logs stored in XML database. (see Section VI)

*2) Checkpoint Handler Layer:* This component is added to provided ability of rebuiding Node Abstration component When checkpoints are loaded by Replay Controller Layer.

## IV. NETWORK AND PACKET SPECIFICATION

One of the most prominent characteristics of NViz is flexibility. To achieve this, NViz allows users to characterize their mote platform as well as packet formats through two XML files: Network Specification and Packet Specification.

### A. Network Specification

Nowadays, there are many mote platforms available on the market. Each platform is designed and optimized for a particular application. Therefore, each node has some metadata that is constant for the entire deployment lifetime such as: set of sensors, configuration, conversion functions… Moreover, each application scenario also needs to consider specific attributes of link or environment. In NViz, all these information are specified by a network specification file and fixed during network operation. Fig. 4 shows how a WSN can be specified in NViz:

```xml
<NetworkSpecification LogPerCheckpoint="100">
    <NodeProperties>
        <Property ID="0" convert="x" length="2"
                  max="65535" min="0">Address</Property>
        <Property ID="1" convert="x" length="1"
                  max="4" min="1">Function</Property>
        <Property ID="2" convert="x*122.3/[Vref]" length="2"
                  max="1023" min="0">Temperature</Property>
        <Property ID="3" convert="1.223*1024/x" length="2"
                  max="1023" min="0">Vref</Property>
    </NodeProperties>
    <LinkProperties>
        <Property ID="1" convert="x" length="1"
                  max="255" min="1">Strength</Property>
    </LinkProperties>
    <EnvrProperties>
        <Property ID="1" convert="x" length="1"
                  max="23" min="0">Channel</Property>
        <Property ID="2" convert="x" length="2"
                  max="65535" min="0">PANID</Property>
    </EnvrProperties>
</NetworkSpecification>
```

Figure 4. An example Network Specification XML file

Network in NViz is specified by three types of property consisting of node, link and environment property. Each property possesses the following attributes:

- *ID*: Each property has a unique ID number different from all other properties of the same type. This ID will be used in Packet Specification file to determine what property is associated with particular packet's field.

- *Convert*: Conversion function of property.

- *Length*: Number of bytes of property. This information is also used to determine length of packet's field associated with this property.

- *Max, Min*: Maximum and minimum values of property (before being converted). These attributes will be used to validate incoming packets.

- *Name*: Name of property.

### B. Packet Specification

To be able to work with any kind of application scenario, NViz allows user to define their packet formats in a packet specification file. Fig. 5 represents an example of packet specification file.

```xml
<PacketSpecification PacketIDLength="2">
    <Packet ID="1" description="Associate">
        <Field ID="0" type="Node">SourceAddress</Field>
        <Field ID="0" type="Node">DestinationAddress</Field>
        <Field ID="1" type="Link">LinkStrength</Field>
        <Field ID="1" type="Node">NodeFunction</Field>
    </Packet>
    <Packet ID="2" description="UpdateTemparature">
        <Field ID="0" type="Node">NodeAddress</Field>
        <Field ID="3" type="Node">VRef</Field>
        <Field ID="2" type="Node">Temperature</Field>
    </Packet>
    <Packet ID="3" description="Update Function">
        <Field ID="0" type="Node">NodeAddress</Field>
        <Field ID="1" type="Node">Function</Field>
    </Packet>
    <Packet ID="4" description="Update Enviroment">
        <Field ID="1" type="Envr">Channel</Field>
        <Field ID="2" type="Envr">PANID</Field>
    </Packet>
</PacketSpecification>
```

Figure 5. An example Packet Specification XML file

In NViz, every packet starts with packet ID field which determines its packet type. Length of packet ID number is set through attribute PacketIDLength of PacketSpecification element. Moreover, each packet type also has description attribute which defines its name. Each packet consists of a number of fields. Each field is specified by its associated property and its name. Associated property is determined by its ID and type.

NViz also provides wizards to help user define their Network Specification and Packet Specification files easily.

## V. UNIT CONVERSION

Unit conversion layer takes responsibility of combining raw sensor data and calibration coefficients to return final data in engineering units for Visualization Abstraction layer. Moreover, conversion functions are defined in Network Specification file and can be changed easily for particular application.

Every property in NViz has a unit conversation function which has a general form:

$$y = F(x, c_0, …, c_n, d_0, …, d_n)$$

Where **x** is the raw sensing data of current property, **ci** is the set of calibration coefficients and **di** is the set of dependant variables. Dependant variables are other properties which has the same type of current property.

An example of using conversion functions in WSNs which uses micaz platform is showed Fig.4. All sensing data in micaz are depended on reference voltage which depends on battery level. Because ADC in micaz is 10-bit, we have following equation:

$$V_{input} = \frac{ADC_{temperature} * V_{ref}}{1024}$$

Assume that we use a linear temperature sensor (like LM35DZ) whose output voltage is linearly proportional to the Celsius temperature (e.g. 10mV/°C) and $V_{ref}$ = 122.3mV. We will have the final conversion function for temperature property:

$$Temperature = \frac{ADC_{temperature} * 122.3}{ADC_{battery}} °C$$

Therefore, in Fig. 4, Vref property has the conversation function: "1.223*1024/x" and Temperature property has the conversation function: "x*122.3/[Vref]"

## VI. REPLAY MECHANISM

NViz implements a new replay mechanism to better support users to observe or debug their networks. When the packets arrive at host, NViz not only displays them on screen but also keeps a record on them. Record in NViz consists of checkpoints. Each checkpoint contains network state information, and a list of logs.

```
<Checkpoint t="1328163686181">
  <Node addr="0" att1="1.0" att2="102.0" att3="378.0">
    <Link att1="123.0" dest="1"/>
    <Link att1="213.0" dest="2"/>
  </Node>
  <Node addr="1" att1="2.0" att2="103.0" att3="356.0">
    <Link att1="158.0" dest="3"/>
    <Link att1="153.0" dest="4"/>
  </Node>
  <Node addr="2" att1="2.0">
    <Link att1="154.0" dest="6"/>
  </Node>
  <Node addr="3" att1="2.0">
    <Link att1="143.0" dest="5"/>
  </Node>
  <Node addr="4" att1="3.0"/>
  <Node addr="5" att1="3.0"/>
  <Node addr="6" att1="3.0"/>
  <Envr att1="11.0" att2="1.0"/>
  <L p="0|2|0|3|1|6F|0|7B|" t="1328163457311"/>
  <L p="0|2|0|4|1|65|0|70|" t="1328163469215"/>
  <L p="0|2|0|5|1|92|0|84|" t="1328163488303"/>
  <L p="0|2|0|6|1|86|0|79|" t="1328163509031"/>
  <L p="0|1|0|2|0|7|91|3|" t="1328163529150"/>
  <L p="0|2|0|7|1|9C|0|80|" t="1328163551462"/>
  <L p="0|2|0|0|1|77|0|68|" t="1328163580910"/>
  <L p="0|2|0|1|1|62|0|66|" t="1328163603094"/>
  <L p="0|2|0|3|1|6D|0|79|" t="1328163625542"/>
  <L p="0|2|0|4|1|63|0|71|" t="1328163646006"/>
</Checkpoint>
```

Figure 7. An example of checkpoint file

Fig. 7 represents an example of checkpoint in practice. Every checkpoint in NViz has "t" attribute which determines when it was created. Network state information is represented in checkpoint by a list of Node, Link and Envr elements. All of this information allow Checkpoint Handler component in NViz-client to reconstruct the network completely. In addition, checkpoint also contains a number of logs which are stored in "L" elements. Number of logs per checkpoint is set by "LogPerCheckpoint" in Network Specification file (Fig. 4). Logs contain information about incoming packets, including receiving time and packets data.

## VII. CONCLUSION

In this paper, a general-purpose visualization tool – NViz, which can represent and visualize data for any kind of WSN, is proposed. This tool allows users to set their network's properties and packet formats through XML files. Based on properties defined, user can choose the meaning of them and let NViz represents the data respectively. Replay mechanism, which lets researchers and developers debug their WSN easily, is also integrated in this tool. Furthermore, by defining conversion functions in Network Specification file (and change easily for particular application), NViz makes sensor data understandable to human beings.

NViz is designed based on a layered architecture which allows for clear and well-defined interrelationships and interfaces between each components. More works in the near future will be carried out to prove the universality and efficiency of the proposed architecture.